\documentclass[journal,12pt,onecolumn,draftclsnofoot,]{IEEEtran}
\usepackage{authblk}
\usepackage[ruled, vlined]{algorithm2e}
\usepackage{algpseudocode}
\usepackage{amsmath}
\usepackage{amssymb}
\usepackage{array}
\usepackage{arydshln}
\usepackage{blkarray, bigstrut}
\usepackage{bm}
\usepackage{booktabs}
\usepackage{braket}
\usepackage{cite}
\usepackage{enumitem}
\usepackage{graphicx}
\usepackage[mathscr]{euscript}
\usepackage{mathtools}
\usepackage{multirow}
\usepackage{framed} 
\usepackage[framed]{ntheorem}
\usepackage{nicefrac}
\usepackage{pgfplots}
\usepackage{qtree}
\usepackage{adjustbox}
\usepackage{rotating}
\pgfplotsset{compat = 1.13}
\usepackage{stfloats}
\usepackage{subfig}
\usepackage{url}
\usepackage{tikz}
\usepackage{tkz-berge}
\usetikzlibrary{patterns}
\usetikzlibrary{spy}
\usepackage[framemethod=TikZ]{mdframed}
\usetikzlibrary{shapes, backgrounds,calc, snakes}
\usetikzlibrary{decorations.pathreplacing}
\usepackage{fancybox}

\tikzstyle{vertex} = [circle, draw, inner sep = 0pt, minimum size = 7pt]
\newcommand{\vertex}{\node[vertex]}
\newframedtheorem{frm-prop}{Property}

\definecolor{bblue}{rgb}{0.11, 0.22, 0.73}
\definecolor{rred}{rgb}{1, 0.0, 0.16}
\definecolor{oorange}{rgb}{1.0, 0.55, 0.0}
\definecolor{ggreen}{rgb}{0.65, 1.0, 0.0}

\definecolor{color0}{HTML}{FF0147}
\definecolor{color1}{HTML}{F400DC}
\definecolor{color2}{HTML}{BA0DFF}
\definecolor{color3}{HTML}{5700E8}
\definecolor{color4}{HTML}{0B03FF}
\definecolor{color5}{HTML}{0957F4}
\definecolor{color6}{HTML}{03B3FF}
\definecolor{color7}{HTML}{08E8DA}
\definecolor{color8}{HTML}{07FF8E}
\definecolor{color9}{HTML}{51FF0A}

\begin{document}

\title{Impact of Quantized Side Information on Subchannel Scheduling for Cellular V2X}

\author[1]{Luis F. Abanto-Leon}
\author[2]{Arie Koppelaar}
\author[1]{Chetan Belagal Math}
\author[1]{Sonia Heemstra de Groot}
\affil[1]{Department of Electrical Engineering, Eindhoven Univerity of Technology}
\affil[2]{NXP Semiconductors, Eindhoven}

\maketitle

\begin{abstract}
	In Release 14, 3GPP completed a first version of cellular vehicle--to--everything (C-V2X) communications wherein two modalities were introduced. One of these schemes, known as \textit{mode-3}, requires support from eNodeBs in order to realize subchannel scheduling. This paper discusses a graph theoretical approach for semi-persistent scheduling (SPS) in \textit{mode-3} harnessing a sensing mechanism whereby vehicles can monitor signal--to--interference--plus--noise ratio (SINR) levels across sidelink subchannels. eNodeBs request such measurements from vehicles and utilize them to accomplish suitable subchannel assignments. However, since SINR values---herein also referred to as side information---span a wide range, quantization is required. We conclude that 3 bits per vehicle every 100 ms can provide sufficient granularity to maintain appropriate performance without severe degradation. Furthermore, the proposed algorithm is compared against pseudo-random and greedy SPS algorithms.
\end{abstract}

\begin{IEEEkeywords}
	cellular V2X, semi-persistent scheduling, safety applications
\end{IEEEkeywords}

\IEEEpeerreviewmaketitle

\section{Introduction}
In the last months we have been witness to an enormous effort from industry and academia in improving the state of the art of vehicular communications, which not only promises to diminish the number of accidents and casualties \cite{b1} but will also pave the way for novel business models and improved services. In September 2017, the 3rd Generation Partnership Project (3GPP) completed a first version of cellular vehicle-to-everything (C-V2X) in Release 14, aiming at principally supporting safety applications \cite{b3}. In the mentioned document, two subchannel scheduling modalities were proposed in order to support communications among vehicles. One of them is vehicle--to--vehicle (V2V) \textit{mode-3}, which requires network support in order to realize subchannel scheduling. The other scheme---known as V2V \textit{mode-4} \cite{b8}---has been devised to operate in a distributed and autonomous manner, which may prove advantageous when cellular network is not available.
 
In \textit{mode-3}, once subchannels have been assigned by the eNodeB, vehicles are able to broadcast cooperative awareness messages (CAMs) \cite{b2} in semi-persistently reserved subchannels. Such utilization of resources extends for several hundreds of milliseconds until a new assignment is required \cite{b3}. In a typical scenario there may exist several communication clusters\footnote{ Clusters are formed based on their likeness in position, speed and direction computed by means of an affinity Gaussian kernel as described in \cite{b9}.} with each of them exhibiting different characteristics, e.g. high or low vehicle density. A most important task for eNodeBs is to guarantee that vehicles---belonging to the same communication cluster---broadcast CAM messages in time-orthogonal subchannels to prevent allocation conflicts \cite{b6}. Such orthogonality requirement is necessary due to half-duplex PHY which cannot support simultaneous transmission and reception. Nevertheless, a subchannel that is serving a vehicle in a particular communication cluster can be reutilized by other vehicles, if the latter ones lie in different clusters. As a result, in \textit{mode-3} operation, eNodeBs will unquestionably play a determinative role in effectively distributing subchannels among vehicles in coverage.

In the present work, a graph theoretical approach for semi-persistent scheduling (SPS) in V2V \textit{mode-3} is proposed. Under this framework, vehicles and subchannels are represented by vertices whereas the achievable rates that vehicles can attain across subchannels are represented by edges. It has been proved that the time orthogonality constraint mentioned earlier can be enforced by aggregating conflicting vertices into macro-vertices \cite{b4}. The approach discussed herein requires side information such as signal--to--interference--plus--noise ratio (SINR) measurements sensed by vehicles across the subchannels. Such quality measurements are passed on by vehicles via uplink for eNodeBs to optimize the assignment of subchannels in terms of the system sum-capacity maximization. Although side information should ideally be fine-grained, it occupies valuable uplink spectrum resources. Therefore, how does quantization of SINR measurements affect the system performance is evaluated for different degrees of granularity.   

The contributions of this work are outlined in the following.
\begin{itemize}
	\item The sensing mechanism on vehicles for \textit{mode-4} is repurposed to support \textit{mode-3} operation. The SINR measurements are transmitted as side information to eNodeBs in order for them to perform subchannel assignment.
	\item Subchannel assignment is approached as a bipartite graph matching problem. 
	\item A matrix notation that includes intra-cluster allocation constraints is developed.
	\item The impact of SINR quantization on the proposed approach performance is assessed and a suitable granularity level is determined.
\end{itemize}

The paper is organized as follows. In Section II, the structure of sidelink subchannels for V2V communications is described. In Section III, the semi-persistent scheduling operation is briefly described. In Section IV, the subchannel allocation problem is formulated. In Section V, the proposed approach based on graph matching is described in detail. Section VI is devoted for discussing simulation results. Finally, in Section VII, our concluding remarks are provided.

\section{Sidelink Subchannels}
\begin{figure*}[!t]
	\centering
	\begin{tikzpicture}
	
	\draw (0,0.0) rectangle (1.5,-0.5);
	\draw (0,-0.5) rectangle (1.5,-1.0);
	\draw (0,-1.5) rectangle (1.5,-2.0);
	\draw (1.5,0.0) rectangle (3,-0.5);
	\draw (1.5,-0.5) rectangle (3,-1.0);
	\draw (1.5,-1.5) rectangle (3,-2.0);
	\draw (3.5,0.0) rectangle (5.0,-0.5);
	\draw (3.5,-0.5) rectangle (5.0,-1.0);
	\draw (3.5,-1.5) rectangle (5.0,-2.0);
	
	\draw (5.0,0.0) rectangle (6.5,-0.5);
	\draw (5.0,-0.5) rectangle (6.5,-1.0);
	\draw (5.0,-1.5) rectangle (6.5,-2.0);
	\draw (6.5,0.0) rectangle (8.0,-0.5);
	\draw (6.5,-0.5) rectangle (8.0,-1.0);
	\draw (6.5,-1.5) rectangle (8.0,-2.0);
	\draw (8.5,0.0) rectangle (10.0,-0.5);
	\draw (8.5,-0.5) rectangle (10.0,-1.0);
	\draw (8.5,-1.5) rectangle (10.0,-2.0);
	
	\draw (10.0,0.0) rectangle (11.5,-0.5);
	\draw (10.0,-0.5) rectangle (11.5,-1.0);
	\draw (10.0,-1.5) rectangle (11.5,-2.0);
	\draw (11.5,0.0) rectangle (13.0,-0.5);
	\draw (11.5,-0.5) rectangle (13.0,-1.0);
	\draw (11.5,-1.5) rectangle (13.0,-2.0);
	\draw (13.5,0.0) rectangle (15.0,-0.5);
	\draw (13.5,-0.5) rectangle (15.0,-1.0);
	\draw (13.5,-1.5) rectangle (15.0,-2.0);
	
	\draw[fill=color1] (0,0) rectangle (1.5,-0.5);
	\draw[fill=color1] (5,0) rectangle (6.5,-0.5);
	\draw[fill=color1] (10,0) rectangle (11.5,-0.5);
	
	\draw[fill=color8] (1.5,-1.5) rectangle (3.0,-2.0);
	\draw[fill=color8] (6.5,-1.5) rectangle (8.0,-2.0);
	\draw[fill=color8] (11.5,-1.5) rectangle (13.0,-2.0);

	\draw[fill=color6] (3.5,-0.5) rectangle (5.0,-1.0);
	\draw[fill=color6] (8.5,-0.5) rectangle (10.0,-1.0);
	\draw[fill=color6] (13.5,-0.5) rectangle (15.0,-1.0);
	
	\node at (3.25,0) {\dots};
	\node at (3.25,-0.5) {\dots};
	\node at (3.25,-1) {\dots};
	\node at (3.25,-1.5) {\dots};
	\node at (3.25,-2) {\dots};
	
	\node at (8.25,0) {\dots};
	\node at (8.25,-0.5) {\dots};
	\node at (8.25,-1) {\dots};
	\node at (8.25,-1.5) {\dots};
	\node at (8.25,-2) {\dots};
	
	\node at (13.25,0) {\dots};
	\node at (13.25,-0.5) {\dots};
	\node at (13.25,-1) {\dots};
	\node at (13.25,-1.5) {\dots};
	\node at (13.25,-2) {\dots};
	
	\node at (15.25,0) {\dots};
	\node at (15.25,-0.5) {\dots};
	\node at (15.25,-1) {\dots};
	\node at (15.25,-1.5) {\dots};
	\node at (15.25,-2) {\dots};

	\node at (0.0, -1.15) {\vdots};
	\node at (1.5, -1.15) {\vdots};
	\node at (3.0, -1.15) {\vdots};
	\node at (3.5, -1.15) {\vdots};
	\node at (6.5, -1.15) {\vdots};
	\node at (8.0, -1.15) {\vdots};
	\node at (8.5, -1.15) {\vdots};
	\node at (11.5, -1.15) {\vdots};
	\node at (13.0, -1.15) {\vdots};
	\node at (13.5, -1.15) {\vdots};
	\node at (15.0, -1.15) {\vdots};
		
	\draw[decoration={brace, raise=5pt},decorate] (1.45, -2.0) -- node[right=6pt] {} (0.05, -2.0);
	\draw[decoration={brace, raise=5pt},decorate] (2.95, -2.0) -- node[right=6pt] {} (1.55, -2.0);
	\draw[decoration={brace, raise=5pt},decorate] (4.95, -2.0) -- node[right=6pt] {} (3.55, -2.0);
	\draw[decoration={brace, raise=5pt},decorate] (6.45, -2.0) -- node[right=6pt] {} (5.05, -2.0);
	\draw[decoration={brace, raise=5pt},decorate] (7.95, -2.0) -- node[right=6pt] {} (6.55, -2.0);
	\draw[decoration={brace, raise=5pt},decorate] (9.95, -2.0) -- node[right=6pt] {} (8.55, -2.0);
	\draw[decoration={brace, raise=5pt},decorate] (11.45, -2.0) -- node[right=6pt] {} (10.05, -2.0);
	\draw[decoration={brace, raise=5pt},decorate] (12.95, -2.0) -- node[right=6pt] {} (11.55, -2.0);
	\draw[decoration={brace, raise=5pt},decorate] (14.95, -2.0) -- node[right=6pt] {} (13.55, -2.0);
	
	\draw[decoration={brace, raise=5pt},decorate] (4.95, -2.5) -- node[right=6pt] {} (0.05, -2.5);
	\node at (2.5,-3.0) {\footnotesize $TL$ ms};
	
	\draw[decoration={brace, raise=5pt},decorate] (9.95, -2.5) -- node[right=6pt] {} (5.05, -2.5);
	\node at (7.5,-3.0) {\footnotesize $TL$ ms};
	
	\draw[decoration={brace, raise=5pt},decorate] (15.05, -2.5) -- node[right=6pt] {} (10.05, -2.5);
	\node at (12.5,-3.0) {\footnotesize $TL$ ms};
	
	\draw[decoration={brace, raise=5pt},decorate] (15.5, -3.0) -- node[right=6pt] {} (0.0, -3.0);
	\node at (7.75,-3.5) {\footnotesize $T_{SPS} \cdot L$ ms};
	
	\draw[decoration={brace, raise=5pt},decorate] (0, -0.45) -- node[right=6pt] {} (0.0, -0.05);
	\draw[decoration={brace, raise=5pt},decorate] (0, -0.95) -- node[right=6pt] {} (0.0, -0.55);
	\draw[decoration={brace, raise=5pt},decorate] (0, -1.95) -- node[right=6pt] {} (0.0, -1.55);
	
	\draw[decoration={brace, raise=5pt},decorate] (-0.6, -2) -- node[right=6pt] {} (-0.6, 0);
	\node[rotate = 90] at (-1.2,-1) {\small Frequency};
	
	\node at (0.75,-2.5) {\footnotesize $T$ ms};
	\node at (2.25,-2.5) {\footnotesize $T$ ms};
	\node at (4.25,-2.5) {\footnotesize $T$ ms};
	\node at (5.75,-2.5) {\footnotesize $T$ ms};
	\node at (7.25,-2.5) {\footnotesize $T$ ms};
	\node at (9.25,-2.5) {\footnotesize $T$ ms};
	\node at (10.75,-2.5) {\footnotesize $T$ ms};
	\node at (12.25,-2.5) {\footnotesize $T$ ms};
	\node at (14.25,-2.5) {\footnotesize $T$ ms};
	
	\node at (0.75,-0.25) {$r_{1}$};
	\node at (0.75,-0.75) {$r_{2}$};
	\node at (0.75,-1.75) {$r_{K}$};
	\node at (2.25,-0.25) {$r_{K+1}$};
	\node at (2.25,-0.75) {$r_{K+2}$};
	\node at (2.25,-1.75) {$r_{2K}$};
	\node at (4.25,-0.25) {\small $r_{K(L-1) + 1}$};
	\node at (4.25,-0.75) {\small $r_{K(L-1) + 2}$};
	\node at (4.25,-1.75) {$r_{KL}$};
	
	\node at (5.75,-0.25) {$r_{1}$};
	\node at (5.75,-0.75) {$r_{2}$};
	\node at (5.75,-1.75) {$r_{K}$};
	\node at (7.25,-0.25) {$r_{K+1}$};
	\node at (7.25,-0.75) {$r_{K+2}$};
	\node at (7.25,-1.75) {$r_{2K}$};
	\node at (9.25,-0.25) {\small $r_{K(L-1) + 1}$};
	\node at (9.25,-0.75) {\small $r_{K(L-1) + 2}$};
	\node at (9.25,-1.75) {$r_{KL}$};
	
	\node at (10.75,-0.25) {$r_{1}$};
	\node at (10.75,-0.75) {$r_{2}$};
	\node at (10.75,-1.75) {$r_{K}$};
	\node at (12.25,-0.25) {$r_{K+1}$};
	\node at (12.25,-0.75) {$r_{K+2}$};
	\node at (12.25,-1.75) {$r_{2K}$};
	\node at (14.25,-0.25) {\small $r_{K(L-1) + 1}$};
	\node at (14.25,-0.75) {\small $r_{K(L-1) + 2}$};
	\node at (14.25,-1.75) {$r_{KL}$};

	\node[rotate = 90] at (-0.6,-0.25) {$B$};
	\node[rotate = 90] at (-0.6,-0.75) {$B$};
	\node[rotate = 90] at (-0.6,-1.75) {$B$};
	
	\draw[very thick] (5.0,0.3) -- (5.0,-2.3);
	\draw[very thick] (10,0.3) -- (10,-2.3);
	\draw[very thick] (15,0.3) -- (15,-2.3);
	
	\end{tikzpicture}	
	\caption{Semi-persistent scheduling for sidelink subchannels in V2V communications}
	\label{f1}
\end{figure*}
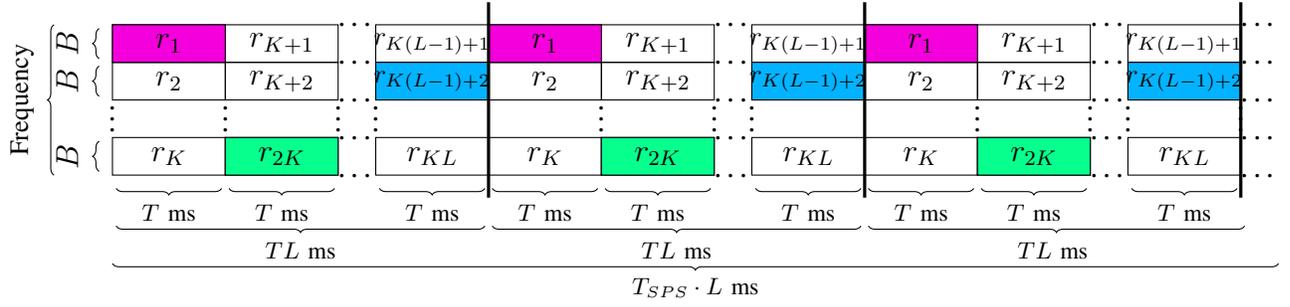

In our system model, it has been considered that spectrum resources for uplink/downlink and sidelink are decoupled. Thus, resources used for V2V communications lie in the intelligent transportation system (ITS) portion \cite{b2} whereas uplink/downlink resources are situated in a different band. The channelization of sidelink spectrum resources can be regarded as a time-frequency grid of non-overlapping subchannels as shown in Fig. \ref{f1}. Each subchannel has an extent of $T$ ms in time and $B$ MHz in frequency. For an observation time window of $TL$ ms, there are $KL$ subchannels spanning $L$ time instances. Furthermore, such distribution of subchannels is repeated over time every $TL$ ms. Each individual subchannel $r_k$ (for $ k = 1, 2, \dots, K $) consists of a number of resource blocks (RBs) for data and control.

\section{Semi-persistent Scheduling}
Once the eNodeB has completed the allocation of subchannels, vehicles will persistently use the designated resources for periodic CAM broadcasting during $T_{SPS}$ ms, as depicted in Fig. \ref{f1}. After the reservation time countdown has reached zero---i.e. when $ T_{SPS} / T $ contiguous time windows have elapsed---a new assignment shall be realized by the eNodeB. The amount of time that vehicles can posses a subchannel may vary for every particular unit. Based on Release 14 \cite{b3}, such time extents (in seconds) can be drawn from $T_{SPS}^{pool} = \{ 1, 4, 8 \}$. When the scenario is not dense, the distribution of subchannels among vehicles is uncomplicated and therefore greedy approaches will usually perform adequately. However, as the scenario becomes congested, the assignment might result more challenging. As a consequence, a problem formulation that can facilitate effective subchannel allocation while complying the system constraints is required.  

\section{Problem Formulation}
Let $J$ denote the total number of communication clusters in the scenario. Each cluster can be denoted as a set $\mathcal{V}^{(j)}$ of vehicles (for $j = 1, 2, \dots, J$). In addition, there exists a set of allotable subchannels $\mathcal{R}$ which are managed by the eNodeB. A different number of vehicles $N_j$ populates each cluster $\mathcal{V}^{(j)}$ and all of them can share the same set of subchannels. In this work, we consider independent clusters that do not overlap, i.e. a vehicle belongs exclusively to only one cluster. The pairing of subchannels with vehicles can be represented as a bipartite graph as depicted in Fig. \ref{f2}, where vehicles and subchannels are portrayed by black and white vertices, respectively. The line linking two vertices---i.e. a vehicle $v^{(j)}_i \in \mathcal{V}^{(j)}$ with a subchannel $r_k \in \mathcal{R}$---is called an edge, and is denoted by $e^{(j)}_{ik}$. Each edge $e^{(j)}_{ik}$ has a corresponding weight $c^{(j)}_{ik}$ that in principle can represent any metric related to the conditions that vehicle $v^{(j)}_i$ experiences in subchannel $r_k$. In this paper, we consider that such weights are the achievable rates obtained from the SINR measurements that vehicles sense in the subchannels. Thus, $c^{(j)}_{ik} = B \log_2(1 + \mathsf{SINR}^{(j)}_{ik})$ represents the achievable rate of vehicle $v^{(j)}_i$ in subchannel $r_k$\footnote{In a more formal manner, $c_{ik}^{(j)}$ and $\mathsf{SINR}_{ik}^{(j)}$ represent measurements between vehicle $v_i^{(j)}$ and some vehicle $v_u^{(j)}$ with which $v_i^{(j)}$ experiences the weakest link quality. Therefore, if the weakest link can be leveraged, other vehicles receiving signals from $v_i^{(j)}$ may experience superior conditions. For the purpose of simplification, the index $u$ representing vehicle $v_u^{(j)}$ has been dropped. The numerator of the SINR is constituted by the received power of interest $P_{T} g_i^u$, where $P_T = 23$ dBm represents the transmit power per CAM message and $g_i^u$ depicts the channel gain between vehicles $v^{(j)}_i$ and $v^{(j)}_u$. In the denominator, besides a noise power component, the influence of interferers is considered as well. The interference is sensed by each vehicle on subchannels belonging to subframes where transmission does not take place.}. The set $\mathcal{R}$ of subchannels is constituted by $KL$ vertices, which are grouped into $L$ disjoint vertex subsets $\{\mathcal{R}_{l}\}_{l = 1}^L$ (herein called macro-vertices) satisfying $\mathcal{R} = \cup_{l = 1}^{L} \mathcal{R}_{l}$ and $\mathcal{R}_{l} \cap \mathcal{R}_{l'} = \emptyset$, $\forall l \neq l'$. Each macro-vertex $\mathcal{R}_{l}$ is an aggregation of $K$ vertices ($\vert\mathcal{R}_{l}\vert = K$), i.e. a collection of subchannels in the same time subframe. The target function is to maximize the sum-rate capacity of the whole system subject to satisfying a set of allocation constraints. Such constraints are ($i$) \textit{the intra-cluster allocation restrictions}---which prevent time-domain conflicts caused due to half-duplex PHY limitations---and ($ii$) \textit{the one--to--one vertex matching conditions}---which enforce each vehicle to be matched with exactly one subchannel. Since we consider that clusters do not overlap with each other, then the problem is equivalent to independently finding a solution vector ${\bf x}_j$---for each cluster $\mathcal{V}^{(j)}$---that maximizes (1a) while satisfying the set of constraints (1b)
\begin{figure}[!t]
	\centering
	\begin{tikzpicture}[scale = 0.72]
	
	\vertex[fill] (v1) at (0,-1.25) [label = left:$v^{(j)}_{1}$] {};
	\vertex[fill] (v2) at (0,-3.25) [label = left:$v^{(j)}_{2}$] {};
	\node at (0,-4.5) {\vdots};
	\vertex[fill] (v3) at (0,-5.75) [label = left:$v^{(j)}_{N_j}$] {};
	
	\vertex (r11) at (4,-0.6) [label = right:$r_{1}$] {};
	\vertex (r12) at (4,-1) [label = right:$r_{2}$] {};
	\node at (4,-1.4) {\vdots};
	\vertex (r13) at (4,-1.9) [label = right:$r_{K}$] {};
	
	\vertex (r21) at (4,-2.6) [label = right:$r_{K+1}$] {};
	\vertex (r22) at (4,-3) [label = right:$r_{K+2}$] {};
	\node at (4,-3.4) {\vdots};
	\vertex (r23) at (4,-3.9) [label = right:$r_{2K}$] {};
	
	\node at (4,-4.4) {\vdots};
	
	\vertex (r31) at (4,-5.1) [label = right:$r_{K(L-1)+1}$] {};
	\vertex (r32) at (4,-5.5) [label =right:$r_{K(L-1)+2}$] {};
	\node at (4,-5.9) {\vdots};
	\vertex (r33) at (4,-6.4) [label = right:$r_{KL}$] {};
	
	\path
	(v1) edge (3.825,-0.6)
	(v1) edge (3.825,-1)
	(v1) edge (3.825,-1.9)
	(v1) edge (3.825,-2.6)
	(v1) edge (3.825,-3)
	(v1) edge (3.825,-3.9)
	(v1) edge (3.825,-5.1)
	(v1) edge (3.825,-5.5)
	(v1) edge (3.825,-6.4)
	
	(v2) edge (3.825,-0.6)
	(v2) edge (3.825,-1)
	(v2) edge (3.825,-1.9)
	(v2) edge (3.825,-2.6)
	(v2) edge (3.825,-3)
	(v2) edge (3.825,-3.9)
	(v2) edge (3.825,-5.1)
	(v2) edge (3.825,-5.5)
	(v2) edge (3.825,-6.4)
	
	(v3) edge (3.825,-0.6)
	(v3) edge (3.825,-1)
	(v3) edge (3.825,-1.9)
	(v3) edge (3.825,-2.6)
	(v3) edge (3.825,-3)
	(v3) edge (3.825,-3.9)
	(v3) edge (3.825,-5.1)
	(v3) edge (3.825,-5.5)
	(v3) edge (3.825,-6.4);
	
	\draw[densely dashed,rounded corners=4]($(r11)+(-.25,.25)$)rectangle($(r13)+(0.25,-.25)$);
	\draw[densely dashed,rounded corners=4]($(r21)+(-.25,.25)$)rectangle($(r23)+(0.25,-.25)$);
	\draw[densely dashed,rounded corners=4]($(r31)+(-.25,.25)$)rectangle($(r33)+(0.25,-.25)$);
	
	\draw[decoration={brace, raise=5pt},decorate] (6.4,-0.4) -- node[right=6pt] {} (6.4,-2.1);
	\draw[decoration={brace, raise=5pt},decorate] (6.4,-2.4) -- node[right=6pt] {} (6.4,-4.1);
	\draw[decoration={brace, raise=5pt},decorate] (6.4,-4.9) -- node[right=6pt] {} (6.4,-6.6);
	
	\node[rotate=-90] at (7.4,-1.25) {macro-};
	\node[rotate=-90] at (7.4,-3.25) {macro-};
	\node[rotate=-90] at (7.4,-5.75) {macro-};
	\node[rotate=-90] at (7,-1.25) {vertex $\mathcal{R}_1$};
	\node[rotate=-90] at (7,-3.25) {vertex $\mathcal{R}_2$};
	\node[rotate=-90] at (7,-5.75) {vertex $\mathcal{R}_L$};
	
	\node[text width = 0.2cm] at (0,-7.0) {$\mathcal{V}^{(j)}$};
	\node[text width = 0.2cm] at (-0.5,-7.5) {\textit{Vehicles}};
	\node[text width = 0.2cm] at (4,-7.0) {$\mathcal{R}$};
	\node[text width = 0.2cm] at (3.4,-7.5) {\textit{Resources}};
	
	\end{tikzpicture}
	\caption{Constrained weighted bipartite graph}
	\label{f2}
\end{figure}
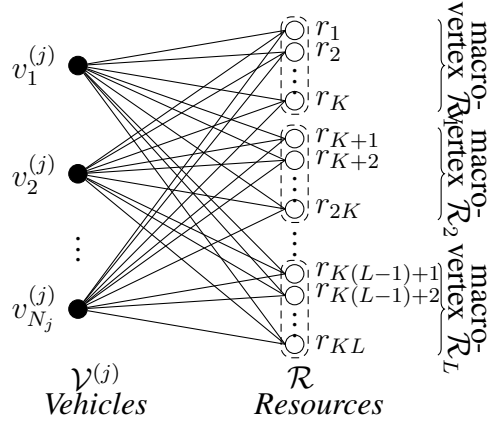
\begin{subequations} \label{e1}
	\begin{gather} 
	\begin{align}
	& {\rm max} ~ {\bf c}^T_j {\bf x}_j \\
	& {\rm subject~to}~ 
	{\left( {\left[
		\begin{array}{c}
		{\bf I}_{L \times L} \otimes {\bf 1}_{1 \times L}\\
		\hline
		{\bf 1}_{1 \times L} \otimes {\bf I}_{L \times L} 
		\end{array}
		\right]} \otimes {\bf 1}_{1 \times K} \right)} {\bf x}_j = {\bf 1}
	\end{align}
	\end{gather}
\end{subequations}
where $\otimes$ represents the tensor product operator, ${\bf {c}}_j \in \mathbb{R}^{M}, {\bf x}_j \in \mathbb{B}^M$ with $M = KL^2$. Note that for graph completeness, dummy vehicles have been added such that $\vert \mathcal{V}^{(j)} \vert = N_j = L$. Therefore, $\mathbf{x}_j = [x^{(j)}_{1,1}, \dots, x^{(j)}_{1,KL}, \dots, x^{(j)}_{L,1}, \dots, x^{(j)}_{L,KL}]^T$ and $\mathbf{c}_j = [c^{(j)}_{1,1}, \dots, c^{(j)}_{1,KL}, \dots, c^{(j)}_{L,1}, \dots, c^{(j)}_{L,KL}]^T$.

\section{Proposed Algorithm}
In any cluster $\mathcal{V}^{(j)}$, the target is to find a vertex--to--vertex matching that maximizes the sum of weights $c^{(j)}_{ik}$ such that no two vertices in $\mathcal{V}^{(j)}$ are matched with any two vertices in the same macro-vertex $\mathcal{R}_l$. This is equivalent to restricting each vehicle $v^{(j)}_i \in \mathcal{V}^{(j)}$ to be assigned a subchannel such that no two vehicles have subchannels in the same time instance (subframe). Moreover, such resultant subchannel assignment must provide the maximum sum-capacity. It was mentioned that time-orthogonality of subchannels is an important requirement within each communication cluster as it prevents conflicts due to half-duplex PHY. Enforcing orthogonality of subchannels in time domain is equivalent to aggregating vertices into macro-vertices \cite{b4}. Thus, (\ref{e1}) can be reduced as follows.

\vspace{0.25cm}
\underline{Simplification of cost function (1a):} \\
\vspace{-0.25cm} \\
The solution vector ${\bf x}_j$ bears information about the matching between subchannels and vehicles. Such designation is represented with 1 or 0. Since ${\bf x}_j$ exists in the binary space, the cost function (1a) can be recast as ${\bf c}^T_j {\bf x}_j = {\bf x}^T_j diag({\bf c}_j) {\bf x}_j$ without altering optimality. Moreover, the weighted pair-wise product $c^{(j)}_{ik} x^{(j)}_{ik} x^{(j)}_{iq}$ (with $r_k, r_q \in \mathcal{R}_{l}, \forall k \neq q $ ) yields zero because no two vehicles can be assigned subchannels that lie in the same time instance. Hence, it can be shown that the sum of weighted pair-wise products will also return zero. In a vectorized form this is equivalent to ${\bf x}^T_j \big( {\bf I}_{M \times M} \otimes [{\bf 1}_{K \times K} - {\bf I}_{K \times K}] \big) diag({\bf c}_j) {\bf x}_j = 0$. 

\begin{mdframed}[] 
	The cost function (1a) can be recast as
	\begin{equation} \label{e2}
	\small
	\hspace{0.1cm}
	\begin{array}{l}
	~ {\bf c}^T_j {\bf x}_j \\
	
	\vspace{-0.2cm} \\
	
	= {\bf x}^T_j  diag({\bf c}_j) {\bf x}_j + \underbrace{{\bf x}^T_j ({\bf I}_{M \times M} \otimes [{\bf 1} - {\bf I}]_{K \times K}) ~  diag({\bf c}_j) {\bf x}_j}_{0} \\
	
	\vspace{-0.2cm} \\
	
	= {\bf x}^T_j ({\bf I}_{M \times M} \otimes {\bf I}_{K \times K} + {\bf I}_{M \times M} \otimes [{\bf 1} - {\bf I}]_{K \times K})  diag({\bf c}_j) {\bf x}_j\\
	
	\vspace{-0.2cm} \\
	
	= {\bf x}^T_j ({\bf I}_{M \times M} \otimes {\bf 1}_{K \times K})  diag({\bf c}_j) {\bf x}_j\\
	
	\end{array}
	\end{equation}
	
	Employing the property of product of tensor products, (\ref{e2}) can be further simplified to
	\begin{equation} \label{e3}
	\hspace{0.1cm}
	\small
	\begin{array}{l}
	~ {\bf x}^T_j ({\bf I}_{M \times M} \otimes {\bf 1}_{K \times K}) diag({\bf c}_j) {\bf x}_j\\
	
	\vspace{-0.2cm} \\
	
	= {\bf x}^T_j ({\bf I}_{M \times M} {\bf I}_{M \times M} \otimes {\bf 1}_{K \times 1} {\bf 1}_{1 \times K}) diag({\bf c}_j) {\bf x}_j\\
	
	\vspace{-0.2cm} \\
	
	= \underbrace{{\bf x}^T_j ({\bf I}_{M \times M} \otimes {\bf 1}_{K \times 1})}_{\text{\bf y}^T_j} 
	\underbrace{({\bf I}_{M \times M} \otimes {\bf 1}_{1 \times K}) diag({\bf c}_j) {\bf x}_j}_{\text{\bf d}_j} \\
	
	\end{array}
	\end{equation}
\end{mdframed}
\vspace{0.15cm}
\underline{Simplification of constraints (1b):} \\
\vspace{-0.50cm} \\
\begin{mdframed}[] 
	The constraints (1b) can be reduced to
	\begin{equation} \label{e4}
	\hspace{0.cm}
	\small
	\begin{array}{l}
	
	~\left( \left[
	\begin{array}{c}
	{\bf I}_{L \times L} \otimes {\bf 1}_{1 \times L}\\
	\hline
	{\bf 1}_{1 \times L} \otimes {\bf I}_{L \times L} 
	\end{array}
	\right] 
	\otimes {\bf 1}_{1 \times K} \right) {\left( {\bf I}_{M \times M} \otimes {\bf 1}_{1 \times K}^{\dagger} \right){\bf y}_j} = {\bf 1}
	
	\vspace{0.3cm} \\
	
	= \left( \left[
	\begin{array}{c}
	{\bf I}_{L \times L} \otimes {\bf 1}_{1 \times L}\\
	\hline
	{\bf 1}_{1 \times L} \otimes {\bf I}_{L \times L} 
	\end{array}
	\right] 
	{\bf I}_{M \times M} \right) \otimes \underbrace{{\left( {\bf 1}_{1 \times K} {\bf 1}_{1 \times K}^{\dagger} \right)}}_\text{1} {\bf y}_j = {\bf 1}
	
	\vspace{0.01cm} \\
	
	= \left[
	\begin{array}{c}
	{\bf I}_{L \times L} \otimes {\bf 1}_{1 \times L}\\
	\hline
	{\bf 1}_{1 \times L} \otimes {\bf I}_{L \times L} 
	\end{array}
	\right] 
	{\bf y}_j = {\bf 1}
	
	\end{array}
	\end{equation}
\end{mdframed}

The formulation in (\ref{e1}) can be recast as (\ref{e5})
\begin{equation} \label{e5}
	\hspace{-1.5cm}
	\begin{array}{lclcl}
	&& {\rm max} ~ {\bf d}^T_j {\bf y}_j \\
	&& {\rm subject~to}~ 
		{\left[
			\begin{array}{c}
			{\bf I}_{L \times L} \otimes {\bf 1}_{1 \times L}\\
			\hline
			{\bf 1}_{1 \times L} \otimes {\bf I}_{L \times L} 
			\end{array}
			\right]} {\bf y}_j = {\bf 1}
	\end{array}
\end{equation}
which exhibits a dimensionality reduction of ${\bf x}_j$ by $K$ times and can intuitively be understood as a vertex compression. Note that ${\bf x}_j$ is unknown and ${\bf d}_j$ depends on it, but such dependency can be removed using (\ref{e8}) 
\begin{equation} \label{e8}
{\bf d}_j = \lim_{\beta \to \infty} \frac{1}{\beta} \overset{\substack{\circ}}{\log} \Big\{({\bf I}_{M \times M}\otimes {\bf 1}_{1 \times K}) \mathrm{e}^{\circ \beta {\bf c}_j} \Big\}
\end{equation}
where $\overset{\substack{\circ}}{\log} \{\cdot\}$ and $\mathrm{e}^{\circ \{ \cdot \} }$ are the element-wise natural logarithm and Hadamard exponential \cite{b7}, respectively.
Now, (\ref{e5}) can be approached by the Kuhn Munkres algorithm \cite{b5}. Conversely, (\ref{e1}) is not approachable by the aforementioned method.
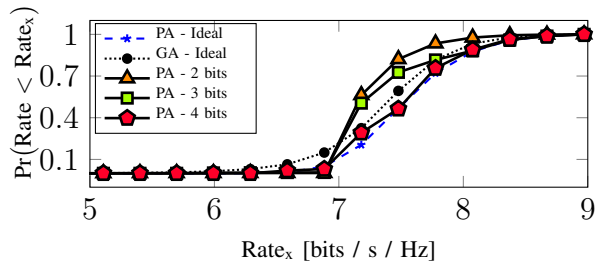
\begin{figure}[!t]
	\centering
	\begin{tikzpicture}
	\begin{axis}[
	xmin = 5,
	xmax = 9,
	width = 8.2cm,
	height = 3.8cm,
	xlabel={Rate\textsubscript{x} [bits / s / Hz]},
	x label style={align=center, font=\footnotesize,},
	ylabel = {Pr\big(Rate~\textless~Rate\textsubscript{x}\big)},
	y label style={at={(-0.08,0.5)}, text width = 2cm, align=center, font=\footnotesize,},
	ytick = {0.1, 0.4, 0.7, 1.0},
	legend style={at={(0.01,0.32)},anchor=south west, font=\fontsize{6}{5}\selectfont, text width=1.15cm,text height=0.06cm,text depth=.ex, fill = none, align = left},
	]
	
	\addplot[color=blue, mark = star, mark options = {scale = 0.8}, line width = 0.8pt, style = dashed] coordinates 
	{
		(1.0581,0.0000)
		(1.2331,0.0000)
		(1.4244,0.0000)
		(1.6310,0.0000)
		(1.8516,0.0000)
		(2.0848,0.0000)
		(2.3290,0.0000)
		(2.5828,0.0000)
		(2.8447,0.0000)
		(3.1135,0.0000)
		(3.3880,0.0000)
		(3.6671,0.0000)
		(3.9502,0.0000)
		(4.2363,0.0000)
		(4.5250,0.0000)
		(4.8158,0.0000)
		(5.1082,0.0000)
		(5.4019,0.0000)
		(5.6967,0.0000)
		(5.9924,0.0000)
		(6.2887,0.0002)
		(6.5856,0.0053)
		(6.8829,0.0492)
		(7.1805,0.2050)
		(7.4784,0.4676)
		(7.7766,0.7211)
		(8.0749,0.8762)
		(8.3733,0.9580)
		(8.6719,0.9870)
		(8.9705,1.0000)
	}; \addlegendentry{PA - Ideal}
	
	\addplot[color=black, mark = *, mark options = {solid, scale = 0.8}, line width = 0.8pt, style = densely dotted] coordinates 
	{
		(1.0581,0.0000)
		(1.2331,0.0000)
		(1.4244,0.0000)
		(1.6310,0.0000)
		(1.8516,0.0000)
		(2.0848,0.0000)
		(2.3290,0.0000)
		(2.5828,0.0000)
		(2.8447,0.0000)
		(3.1135,0.0000)
		(3.3880,0.0000)
		(3.6671,0.0000)
		(3.9502,0.0000)
		(4.2363,0.0000)
		(4.5250,0.0001)
		(4.8158,0.0003)
		(5.1082,0.0015)
		(5.4019,0.0035)
		(5.6967,0.0069)
		(5.9924,0.0146)
		(6.2887,0.0303)
		(6.5856,0.0645)
		(6.8829,0.1493)
		(7.1805,0.3248)
		(7.4784,0.5931)
		(7.7766,0.8169)
		(8.0749,0.9329)
		(8.3733,0.9782)
		(8.6719,0.9934)
		(8.9705,1.0000)
	}; \addlegendentry{GA - Ideal}

	\addplot[color=black, mark = triangle*, mark options = {scale = 1.5, fill = oorange}, line width = 1pt] coordinates 
	{
		(1.0581,0.0000)
		(1.2331,0.0000)
		(1.4244,0.0000)
		(1.6310,0.0000)
		(1.8516,0.0000)
		(2.0848,0.0000)
		(2.3290,0.0000)
		(2.5828,0.0000)
		(2.8447,0.0000)
		(3.1135,0.0000)
		(3.3880,0.0000)
		(3.6671,0.0000)
		(3.9502,0.0000)
		(4.2363,0.0000)
		(4.5250,0.0000)
		(4.8158,0.0002)
		(5.1082,0.0005)
		(5.4019,0.0010)
		(5.6967,0.0014)
		(5.9924,0.0016)
		(6.2887,0.0019)
		(6.5856,0.0022)
		(6.8829,0.0022)
		(7.1805,0.5628)
		(7.4784,0.8216)
		(7.7766,0.9347)
		(8.0749,0.9768)
		(8.3733,0.9935)
		(8.6719,0.9982)
		(8.9705,1.0000)
	}; \addlegendentry{PA - 2 bits}

	\addplot[color=black, mark = square*, mark options = {fill = ggreen}, line width = 1pt] coordinates 
	{
		(1.0581,0.0000)
		(1.2331,0.0000)
		(1.4244,0.0000)
		(1.6310,0.0000)
		(1.8516,0.0000)
		(2.0848,0.0000)
		(2.3290,0.0000)
		(2.5828,0.0000)
		(2.8447,0.0000)
		(3.1135,0.0000)
		(3.3880,0.0000)
		(3.6671,0.0000)
		(3.9502,0.0000)
		(4.2363,0.0000)
		(4.5250,0.0000)
		(4.8158,0.0000)
		(5.1082,0.0000)
		(5.4019,0.0000)
		(5.6967,0.0000)
		(5.9924,0.0028)
		(6.2887,0.0044)
		(6.5856,0.0054)
		(6.8829,0.0059)
		(7.1805,0.5051)
		(7.4784,0.7265)
		(7.7766,0.8159)
		(8.0749,0.8826)
		(8.3733,0.9598)
		(8.6719,0.9877)
		(8.9705,1.0000)
	}; \addlegendentry{PA - 3 bits}

	\addplot[color = black, mark = pentagon*, mark options = {scale = 1.5, fill = rred, solid}, line width = 1pt] coordinates 
	{
		(1.0581,0.0000)
		(1.2331,0.0000)
		(1.4244,0.0000)
		(1.6310,0.0000)
		(1.8516,0.0000)
		(2.0848,0.0000)
		(2.3290,0.0000)
		(2.5828,0.0000)
		(2.8447,0.0000)
		(3.1135,0.0000)
		(3.3880,0.0000)
		(3.6671,0.0000)
		(3.9502,0.0000)
		(4.2363,0.0000)
		(4.5250,0.0000)
		(4.8158,0.0000)
		(5.1082,0.0000)
		(5.4019,0.0000)
		(5.6967,0.0000)
		(5.9924,0.0000)
		(6.2887,0.0000)
		(6.5856,0.0194)
		(6.8829,0.0307)
		(7.1805,0.2917)
		(7.4784,0.4641)
		(7.7766,0.7568)
		(8.0749,0.8895)
		(8.3733,0.9646)
		(8.6719,0.9875)
		(8.9705,1.0000)	
	}; \addlegendentry{PA - 4 bits}
	
	\end{axis}
	\end{tikzpicture}
	\caption{CDF function for the proposed approach} 
	\label{f3}
	\vspace{-0.25cm}
\end{figure}
\begin{figure}[!t]
	\centering
	\begin{tikzpicture}
	\begin{axis}[
	xmin = 5,
	xmax = 9,
	width = 8.2cm,
	height = 3.8cm,
	xlabel={Rate\textsubscript{x} [bits / s / Hz]},
	x label style={align=center, font=\footnotesize,},
	ylabel = {Pr\big(Rate~\textless~Rate\textsubscript{x}\big)},
	y label style={at={(-0.08,0.5)}, text width = 2cm, align=center, font=\footnotesize,},
	ytick = {0.1, 0.4, 0.7, 1.0},
	legend style={at={(0.01,0.32)},anchor=south west, font=\fontsize{6}{5}\selectfont, text width=1.15cm,text height=0.06cm,text depth=.ex, fill = none, align = left},
	]
	
	\addplot[color=blue, mark = star, mark options = {scale = 0.8}, line width = 0.8pt, style = dashed] coordinates
	{
		(1.0581,0.0000)
		(1.2331,0.0000)
		(1.4244,0.0000)
		(1.6310,0.0000)
		(1.8516,0.0000)
		(2.0848,0.0000)
		(2.3290,0.0000)
		(2.5828,0.0000)
		(2.8447,0.0000)
		(3.1135,0.0000)
		(3.3880,0.0000)
		(3.6671,0.0000)
		(3.9502,0.0000)
		(4.2363,0.0000)
		(4.5250,0.0000)
		(4.8158,0.0000)
		(5.1082,0.0000)
		(5.4019,0.0000)
		(5.6967,0.0000)
		(5.9924,0.0000)
		(6.2887,0.0002)
		(6.5856,0.0053)
		(6.8829,0.0492)
		(7.1805,0.2050)
		(7.4784,0.4676)
		(7.7766,0.7211)
		(8.0749,0.8762)
		(8.3733,0.9580)
		(8.6719,0.9870)
		(8.9705,1.0000)
	}; \addlegendentry{PA - Ideal}
	
	\addplot[color=black, mark = *, mark options = {solid, scale = 0.8}, line width = 0.8pt, style = densely dotted] coordinates 
	{
		(1.0581,0.0000)
		(1.2331,0.0000)
		(1.4244,0.0000)
		(1.6310,0.0000)
		(1.8516,0.0000)
		(2.0848,0.0000)
		(2.3290,0.0000)
		(2.5828,0.0000)
		(2.8447,0.0000)
		(3.1135,0.0000)
		(3.3880,0.0000)
		(3.6671,0.0000)
		(3.9502,0.0000)
		(4.2363,0.0000)
		(4.5250,0.0001)
		(4.8158,0.0003)
		(5.1082,0.0015)
		(5.4019,0.0035)
		(5.6967,0.0069)
		(5.9924,0.0146)
		(6.2887,0.0303)
		(6.5856,0.0645)
		(6.8829,0.1493)
		(7.1805,0.3248)
		(7.4784,0.5931)
		(7.7766,0.8169)
		(8.0749,0.9329)
		(8.3733,0.9782)
		(8.6719,0.9934)
		(8.9705,1.0000)
	}; \addlegendentry{GA - Ideal}

	\addplot[color=black, mark = triangle*, mark options = {scale = 1.5, fill = oorange}, line width = 1pt] coordinates 
	{
		(1.0581,0.0000)
		(1.2331,0.0000)
		(1.4244,0.0000)
		(1.6310,0.0000)
		(1.8516,0.0000)
		(2.0848,0.0000)
		(2.3290,0.0000)
		(2.5828,0.0000)
		(2.8447,0.0000)
		(3.1135,0.0000)
		(3.3880,0.0000)
		(3.6671,0.0000)
		(3.9502,0.0002)
		(4.2363,0.0002)
		(4.5250,0.0002)
		(4.8158,0.0197)
		(5.1082,0.0532)
		(5.4019,0.0847)
		(5.6967,0.1072)
		(5.9924,0.1237)
		(6.2887,0.1341)
		(6.5856,0.1411)
		(6.8829,0.1445)
		(7.1805,0.6234)
		(7.4784,0.8486)
		(7.7766,0.9460)
		(8.0749,0.9821)
		(8.3733,0.9937)
		(8.6719,0.9983)
		(8.9705,1.0000)
	}; \addlegendentry{GA - 2 bits}

	\addplot[color=black, mark = square*, mark options = {fill = ggreen}, line width = 1pt] coordinates 
	{
		(1.0581,0.0000)
		(1.2331,0.0000)
		(1.4244,0.0000)
		(1.6310,0.0000)
		(1.8516,0.0000)
		(2.0848,0.0000)
		(2.3290,0.0000)
		(2.5828,0.0000)
		(2.8447,0.0000)
		(3.1135,0.0000)
		(3.3880,0.0000)
		(3.6671,0.0001)
		(3.9502,0.0001)
		(4.2363,0.0002)
		(4.5250,0.0004)
		(4.8158,0.0025)
		(5.1082,0.0051)
		(5.4019,0.0076)
		(5.6967,0.0088)
		(5.9924,0.0615)
		(6.2887,0.1068)
		(6.5856,0.1340)
		(6.8829,0.1522)
		(7.1805,0.5919)
		(7.4784,0.7980)
		(7.7766,0.8816)
		(8.0749,0.9342)
		(8.3733,0.9785)
		(8.6719,0.9941)
		(8.9705,1.0000)
	}; \addlegendentry{GA - 3 bits}

	\addplot[color = black, mark = pentagon*, mark options = {scale = 1.5, fill = rred, solid}, line width = 1pt] coordinates 
	{
		(1.0581,0.0000)
		(1.2331,0.0000)
		(1.4244,0.0000)
		(1.6310,0.0000)
		(1.8516,0.0000)
		(2.0848,0.0000)
		(2.3290,0.0000)
		(2.5828,0.0000)
		(2.8447,0.0000)
		(3.1135,0.0000)
		(3.3880,0.0000)
		(3.6671,0.0000)
		(3.9502,0.0000)
		(4.2363,0.0000)
		(4.5250,0.0000)
		(4.8158,0.0000)
		(5.1082,0.0010)
		(5.4019,0.0038)
		(5.6967,0.0063)
		(5.9924,0.0210)
		(6.2887,0.0332)
		(6.5856,0.1014)
		(6.8829,0.1480)
		(7.1805,0.4370)
		(7.4784,0.6057)
		(7.7766,0.8339)
		(8.0749,0.9289)
		(8.3733,0.9786)
		(8.6719,0.9932)
		(8.9705,1.0000)
	}; \addlegendentry{GA - 4 bits}
	
	\end{axis}
	\end{tikzpicture}
	\caption{CDF function for greedy algorithm} 
	\label{f4}
	\vspace{-0.25cm}
\end{figure}
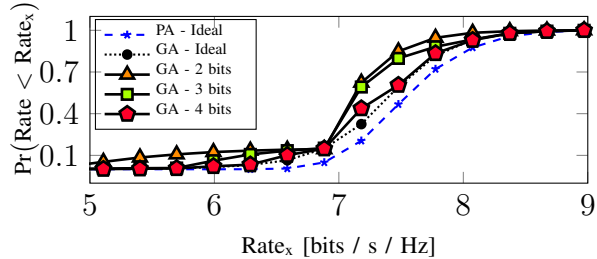

\section{Simulations}
We have considered that each subchannel has a duration of $T = 1$ ms in time and $B = 1.26$ MHz in frequency, which to the best of our understanding is sufficient to support a CAM message and therefore $K = 7$. We also assume a message rate of 10 Hz. Thus, the subchannel assignment is carried out every 0.1 s and therefore $L = 100$. 

Fig. \ref{f3} shows the cumulative distribution function (CDF) of the achievable rates after performing the allocation of subchannels with the proposed approach (PA) when considering $(i)$ ideal fine-grained side information and $(ii)$ three quantization cases\footnote{All the simulations herein have been performed using Matlab programming environment. The vehicular traces have been obtained from the TAPAS Cologne database \cite{b10}  and the range of SINR values encountered approximately spans $[-15; 35]$ dB. Based on this observation, a reasonable amount of quantization bits was selected.}. We observe that when quantization is performed with 4 bits, the corresponding CDF curve behaves similarly to the ideal one. When further reducing the amount of side information to 3 bits and 2 bits, the curve trend is preserved but with some noticeable deviations. In an analogous manner, Fig. \ref{f4} exhibits the CDF  with three different quantization resolutions when the greedy algorithm (GA) is employed. When the SINR values are quantized with 4 bits, the resultant CDF curve is also sufficiently representative of the ideal one as the variation is not substantial. The outcomes based on the proposed approach and the greedy algorithm are both included in the figures, from which it is noted that the former has a performance advantage. The CDF for the random approach (RA) is not shown as for its nature of being stochastic, linear quantization has no impact on it.

In order to gain additional insights regarding a suitable quantization degree, Fig. \ref{f5} and Fig. \ref{f6} throw supplementary information about the examined methods on the basis of four criteria when employing 3 bits and 2 bits, respectively. We can observe that in general the proposed approach attains superior results over the other two schemes. We can notice that the greedy algorithm performs similarly good to the proposed approach if we consider the \textit{highest-rate vehicle} as the comparison criterion. This is an obvious result since the greedy algorithm selects and assigns the best subchannels on first-come first-served basis. Taking the \textit{system average rate} as a judging criterion, the proposed approach has a thin advantage over the greedy algorithm but this may not be reliable metric to asses performance. When considering the \textit{worst-rate vehicle}, the proposed approach outperforms the greedy algorithm for both quantization degrees since---as observed from the results---it is capable of providing a fair data rate to the least favored vehicle. Further examination of the aforementioned criterion reveals that the proposed approach in not strongly influenced by quantization. On the other hand, the greedy algorithm performance is severely affected when quantization is carried out with 2 bits. For instance, when GA employs fine-grained side information, the \textit{worst-rate vehicle} attains 5.75 Mbps. On the other hand, the outcomes for the quantized versions with 3 and 2 bits are 5.47 Mbps and 4.86 Mbps---exhibiting a more noticeable performance gap in the latter case. The last criterion is named \textit{system rate standard deviation} and provides a general insight of how fair an algorithm can be\footnote{The smaller the value is, the fairer the scheme behaves. Thus, a small value depicts quality homogeneity among the alloted subchannels.}. Thus, based on this metric we can notice that the proposed approach behaves better than the greedy algorithm. For all the cases, the random allocation method is surpassed by the other two approaches. 

In order to evaluate how the amount of quantization bits impacts the \textit{worst-rate vehicle}, Fig. \ref{f7}, Fig. \ref{f8} and Fig. \ref{f9} are provided. Such results have been obtained upon employing 2, 3 and 4 quantization bits, respectively. In this scenario, the number of vehicles per cluster has been varied from 10 to 100 whereas the amount of subchannels has remained fixed to 700, i.e, $K = 7$ and $L = 100$. We can observe that as the number of vehicles increases, an efficient distribution of subchannels becomes more challenging. This is particularly true for the greedy algorithm whose performance dramatically drops because the accessibility to subchannels with good conditions gradually depletes when more vehicles populate the system. Nevertheless, the proposed approach is still capable of providing an acceptable level of fairness even when approaching to the overloaded state. When the vehicle density is low, the greedy approach attains high performance because the number of subchannels is larger than the amount of vehicles to be served. However, as the density increases, its performance is affected to a great extent. Judging from such outcomes, if the proposed approach is used, 2 bits of side information are enough as it provides small performance degradation. On the other hand, when the greedy approach is employed, 3 bits are required. As mentioned earlier, the random allocation algorithm suffers almost no variations when the quantization degree is varied. When employing a 1-bit quantization the performance of the proposed approach tends to be similar to the greedy algorithm performance and thus showing no added advantage.

\section{Conclusions}
\begin{figure*}[!t]
	\centering
	\begin{tikzpicture}
	\begin{axis}[
	ybar,
	ymin = 0,
	ymax = 10,
	width = 18.5cm,
	height = 4.5cm,
	bar width = 14pt,
	tick align = inside,
	x label style={align=center, font=\footnotesize,},
	ylabel = {Rate [Mbps / subchannel]},
	y label style={at={(-0.025,0.48)}, font=\footnotesize,},
	nodes near coords,
	every node near coord/.append style={font=\fontsize{6}{5}\selectfont},
	nodes near coords align = {vertical},
	symbolic x coords = {Highest-Rate Vehicle, System Average Rate, Worst-Rate Vehicle, System Rate Standard Deviation},
	x tick label style = {text width = 3.5cm, align = center, font = \footnotesize,},
	xtick = data,
	enlarge y limits = {value = 0.10, upper},
	enlarge x limits = 0.15,
	legend columns=1,
	legend pos = north east,
	legend style={at={(0.75,0.25)},anchor=south west, font=\fontsize{8}{7}\selectfont, text width=3.4cm,text height=0.02cm,text depth=.ex, fill = none, }]
	]
	
	\addplot[fill = rred] coordinates {(Highest-Rate Vehicle, 8.88) (System Average Rate, 7.6888) (Worst-Rate Vehicle, 6.7443) (System Rate Standard Deviation, 0.4323)}; \addlegendentry{Proposed Algorithm}
	
	\addplot[fill = rred, pattern=north east lines] coordinates {(Highest-Rate Vehicle, 8.8649) (System Average Rate, 7.5057) (Worst-Rate Vehicle, 6.6726) (System Rate Standard Deviation, 0.4725)}; \addlegendentry{Proposed Algorithm - 3 bits}
	
	\addplot[fill = oorange] coordinates {(Highest-Rate Vehicle, 8.7901) (System Average Rate, 7.5095) (Worst-Rate Vehicle, 5.7513) (System Rate Standard Deviation, 0.5199)}; \addlegendentry{Greedy Algoritm}
	
	\addplot[fill = oorange, pattern=grid]  coordinates {(Highest-Rate Vehicle, 8.7713) (System Average Rate, 7.2751) (Worst-Rate Vehicle, 5.4694) (System Rate Standard Deviation, 0.5906)}; \addlegendentry{Greedy Algoritm - 3 bits}
	
	\addplot[fill = ggreen] coordinates {(Highest-Rate Vehicle, 7.2852) (System Average Rate, 4.6945) (Worst-Rate Vehicle, 2.3574) (System Rate Standard Deviation, 0.9786)}; \addlegendentry{Random Algorithm}
	
	\addplot[fill = ggreen, pattern=crosshatch dots] coordinates {(Highest-Rate Vehicle, 7.1877) (System Average Rate, 4.6837) (Worst-Rate Vehicle, 2.3401) (System Rate Standard Deviation, 0.9765)}; \addlegendentry{Random Algorithm - 3 bits}
	
	\end{axis}
	\end{tikzpicture}
	\caption{Vehicles data rate: performance comparison between fine-grained vs 3-bit quantization ($N = 100$)}
	\label{f5}
\end{figure*}
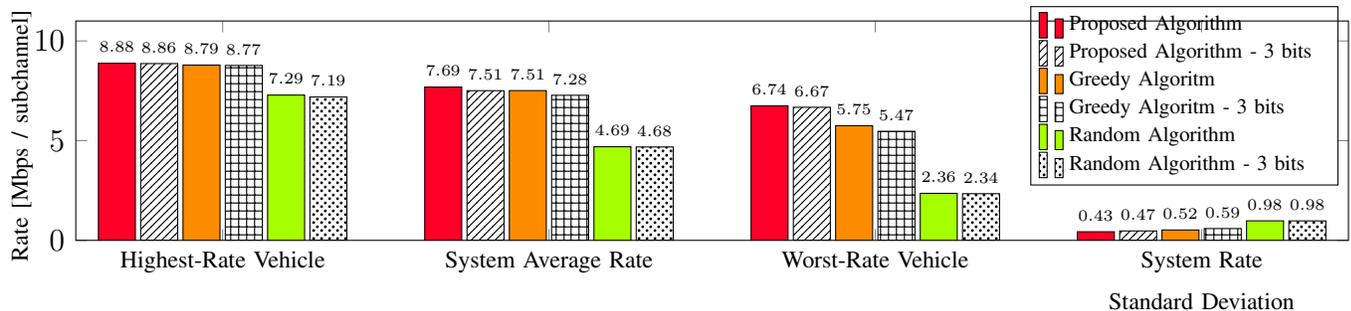
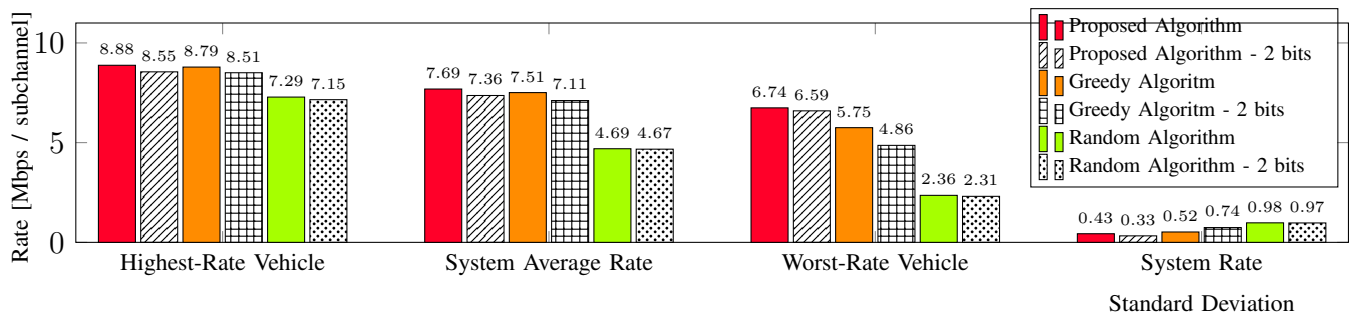
\begin{figure*}[!t]
	\centering
	\begin{tikzpicture}
	\begin{axis}[
	ybar,
	ymin = 0,
	ymax = 10,
	width = 18.5cm,
	height = 4.5cm,
	bar width = 14pt,
	tick align = inside,
	x label style={align=center, font=\footnotesize,},
	ylabel = {Rate [Mbps / subchannel]},
	y label style={at={(-0.025,0.48)}, font=\footnotesize,},
	nodes near coords,
	every node near coord/.append style={font=\fontsize{6}{5}\selectfont},
	nodes near coords align = {vertical},
	symbolic x coords = {Highest-Rate Vehicle, System Average Rate, Worst-Rate Vehicle, System Rate Standard Deviation},
	x tick label style = {text width = 3.5cm, align = center, font = \footnotesize,},
	xtick = data,
	enlarge y limits = {value = 0.10, upper},
	enlarge x limits = 0.15,
	legend columns=1,
	legend pos = north east,
	legend style={at={(0.75,0.25)},anchor=south west, font=\fontsize{8}{7}\selectfont, text width=3.4cm,text height=0.02cm,text depth=.ex, fill = none, }]
	]
	
	\addplot[fill = rred] coordinates {(Highest-Rate Vehicle, 8.88) (System Average Rate, 7.6888) (Worst-Rate Vehicle, 6.7443) (System Rate Standard Deviation, 0.4323)}; \addlegendentry{Proposed Algorithm}
	
	\addplot[fill = rred, pattern=north east lines] coordinates {(Highest-Rate Vehicle, 8.5504) (System Average Rate, 7.3646) (Worst-Rate Vehicle, 6.5943) (System Rate Standard Deviation, 0.3257)}; \addlegendentry{Proposed Algorithm - 2 bits}
	
	\addplot[fill = oorange] coordinates {(Highest-Rate Vehicle, 8.7901) (System Average Rate, 7.5095) (Worst-Rate Vehicle, 5.7513) (System Rate Standard Deviation, 0.5199)}; \addlegendentry{Greedy Algoritm}
	
	\addplot[fill = oorange, pattern=grid] coordinates {(Highest-Rate Vehicle, 8.5057) (System Average Rate, 7.1115) (Worst-Rate Vehicle, 4.8636) (System Rate Standard Deviation, 0.7420)}; \addlegendentry{Greedy Algoritm - 2 bits}
	
	\addplot[fill = ggreen] coordinates {(Highest-Rate Vehicle, 7.2852) (System Average Rate, 4.6945) (Worst-Rate Vehicle, 2.3574) (System Rate Standard Deviation, 0.9786)}; \addlegendentry{Random Algorithm}
	
	\addplot[fill = ggreen, pattern=crosshatch dots] coordinates {(Highest-Rate Vehicle, 7.1547) (System Average Rate, 4.6731) (Worst-Rate Vehicle, 2.3080) (System Rate Standard Deviation, 0.9732)}; \addlegendentry{Random Algorithm - 2 bits}
	
	\end{axis}
	\end{tikzpicture}
	\caption{Vehicles data rate: performance comparison between fine-grained vs 2-bit quantization ($N = 100$)}
	\label{f6}
\end{figure*}
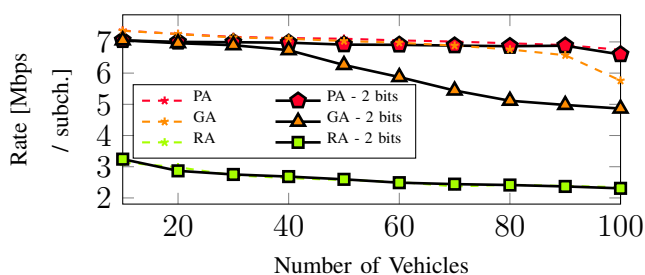
\begin{figure}[!t]
	\centering
	\begin{tikzpicture}
	\begin{axis}[
	xmin = 10,
	xmax = 100,
	width = 8.2cm,
	height = 4.1cm,
	xlabel={Number of Vehicles},
	x label style={align=center, font=\footnotesize,},
	ylabel = {Rate [Mbps / subch.]},
	y label style={at={(-0.08,0.5)}, text width = 2.5cm, align=center, font=\footnotesize,},
	ytick = {2, 3, 4, 5, 6, 7, 8},
	legend columns=2,
	legend style={at={(0.02,0.24)},anchor=south west, font=\fontsize{6}{5}\selectfont, text width=1cm,text height=0.075cm,text depth=.ex, fill = none,},
	]
	
	\addplot[color=rred, mark = star, mark options = {scale = 0.8}, line width = 0.8pt, style = dashed] coordinates 
	{
		(10, 7.3627)
		(20, 7.2525)
		(30, 7.1609)
		(40, 7.1215)
		(50, 7.1014)
		(60, 7.0478)
		(70, 7.0106)
		(80, 6.9548)
		(90, 6.8992)
		(100, 6.7457)
	}; \addlegendentry{PA}

	\addplot[color=black, mark = pentagon*, mark options = {scale = 1.5, fill = rred}, line width = 1pt] coordinates 
	{
		(10, 7.0502)           
		(20, 7.0002)     
		(30, 6.9893)    
		(40, 6.9748)   
		(50, 6.9127)    
		(60, 6.9060)    
		(70, 6.8813)    
		(80, 6.8622)    
		(90, 6.8786)
		(100, 6.5973)    
	}; \addlegendentry{PA - 2 bits}
	
	\addplot[color=oorange, mark = star, mark options = {scale = 0.8}, line width = 0.8pt, style = dashed] coordinates 
	{
		(10, 7.3546)
		(20, 7.2520)
		(30, 7.1420)
		(40, 7.1025)
		(50, 7.0283)
		(60, 6.9738)
		(70, 6.8821)
		(80, 6.7579)
		(90, 6.5726)
		(100, 5.7599)
	}; \addlegendentry{GA}

	\addplot[color=black, mark = triangle*, mark options = {scale = 1.5, fill = oorange}, line width = 1pt] coordinates 
	{
		(10, 7.0497)           
		(20, 6.9588)     
		(30, 6.8941)    
		(40, 6.7353)   
		(50, 6.2604)    
		(60, 5.8733)    
		(70, 5.4418)    
		(80, 5.1145)    
		(90, 4.9795)
		(100, 4.8636)   
	}; \addlegendentry{GA - 2 bits}

	\addplot[color=ggreen, mark = star, mark options = {scale = 0.8}, line width = 0.8pt, style = dashed] coordinates 
{
	(10, 3.1387)
	(20, 2.9843)
	(30, 2.7262)
	(40, 2.6347)
	(50, 2.5755)
	(60, 2.5115)
	(70, 2.3681)
	(80, 2.4221)
	(90, 2.3840)
	(100, 2.3453)
}; \addlegendentry{RA}
	
	\addplot[color=black, mark = square*, mark options = {fill = ggreen}, line width = 1pt] coordinates 
	{
		(10, 3.2432)           
		(20, 2.8678)     
		(30, 2.7522)    
		(40, 2.6845)   
		(50, 2.5962)    
		(60, 2.4888)    
		(70, 2.4422)    
		(80, 2.4137)    
		(90, 2.3685)
		(100, 2.3080) 
	}; \addlegendentry{RA - 2 bits}
	\end{axis}
	\end{tikzpicture}
	\caption{Worst-rate vehicle (2 bits)}
	\label{f7}
	\vspace{-0.25cm}
\end{figure}
\begin{figure}[!t]
	\centering
	\begin{tikzpicture}
	\begin{axis}[
	xmin = 10,
	xmax = 100,
	width = 8.2cm,
	height = 4.1cm,
	xlabel={Number of Vehicles},
	x label style={align=center, font=\footnotesize,},
	ylabel = {Rate [Mbps / subch.]},
	y label style={at={(-0.08,0.5)}, text width = 2.5cm, align=center, font=\footnotesize,},
	ytick = {2, 3, 4, 5, 6, 7, 8},
	legend columns=2,
	legend style={at={(0.02,0.24)},anchor=south west, font=\fontsize{6}{5}\selectfont, text width=1cm,text height=0.075cm,text depth=.ex, fill = none,},
	]
	
	\addplot[color=rred, mark = star, mark options = {scale = 0.8}, line width = 0.8pt, style = dashed] coordinates 
	{
		(10, 7.3627)
		(20, 7.2525)
		(30, 7.1609)
		(40, 7.1215)
		(50, 7.1014)
		(60, 7.0478)
		(70, 7.0106)
		(80, 6.9548)
		(90, 6.8992)
		(100, 6.7457)
	}; \addlegendentry{PA}

	\addplot[color=black, mark = pentagon*, mark options = {scale = 1.5, fill = rred}, line width = 1pt] coordinates 
	{
		(10, 7.0770)           
		(20, 7.0525)     
		(30, 6.9476)    
		(40, 7.0093)   
		(50, 6.9760)    
		(60, 6.9702)    
		(70, 6.8458)    
		(80, 6.9075)    
		(90, 6.9046)
		(100, 6.6487)     
	}; \addlegendentry{PA - 3 bits}

	\addplot[color=oorange, mark = star, mark options = {scale = 0.8}, line width = 0.8pt, style = dashed] coordinates 
	{
		(10, 7.3546)
		(20, 7.2520)
		(30, 7.1420)
		(40, 7.1025)
		(50, 7.0283)
		(60, 6.9738)
		(70, 6.8821)
		(80, 6.7579)
		(90, 6.5726)
		(100, 5.7599)
	}; \addlegendentry{GA}

	\addplot[color=black, mark = triangle*, mark options = {scale = 1.5, fill = oorange}, line width = 1pt] coordinates 
{
	(10, 7.0605)           
	(20, 7.0126)     
	(30, 6.9316)    
	(40, 6.8325)   
	(50, 6.7245)    
	(60, 6.4414)    
	(70, 6.2549)    
	(80, 6.1021)    
	(90, 5.9867)
	(100, 5.5137)  
}; \addlegendentry{GA - 3 bits}
	
	\addplot[color=ggreen, mark = star, mark options = {scale = 0.8}, line width = 0.8pt, style = dashed] coordinates 
	{
		(10, 3.1387)
		(20, 2.9843)
		(30, 2.7262)
		(40, 2.6347)
		(50, 2.5755)
		(60, 2.5115)
		(70, 2.3681)
		(80, 2.4221)
		(90, 2.3840)
		(100, 2.3453)
	}; \addlegendentry{RA}
	
	\addplot[color=black, mark = square*, mark options = {fill = ggreen}, line width = 1pt] coordinates 
	{
		(10, 3.1060)           
		(20, 2.8317)     
		(30, 2.7512)    
		(40, 2.6561)   
		(50, 2.6022)    
		(60, 2.4641)    
		(70, 2.4969)    
		(80, 2.3787)    
		(90, 2.3263)
		(100, 2.2785)
	}; \addlegendentry{RA - 3 bits}
	\end{axis}
	\end{tikzpicture}
	\caption{Worst-rate vehicle (3 bits)}
	\label{f8}
	\vspace{-0.25cm}
\end{figure}
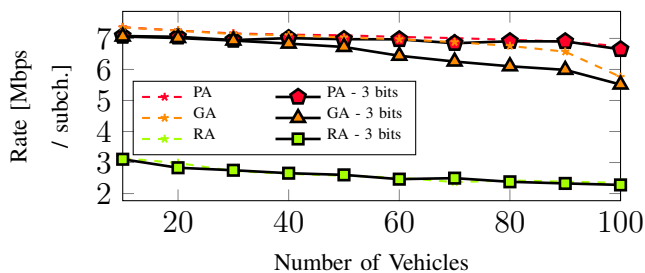
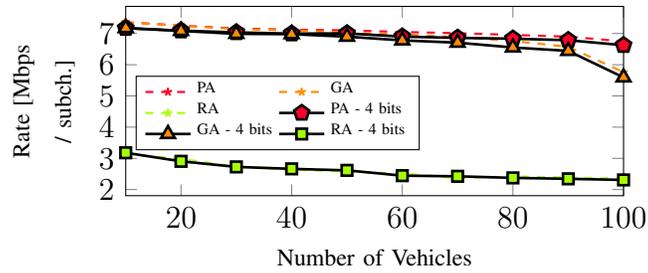
\begin{figure}[!t]
	\centering
	\begin{tikzpicture}
	\begin{axis}[
	xmin = 10,
	xmax = 100,
	width = 8.2cm,
	height = 4.1cm,
	xlabel={Number of Vehicles},
	x label style={align=center, font=\footnotesize,},
	ylabel = {Rate [Mbps / subch.]},
	y label style={at={(-0.08,0.5)}, text width = 2.5cm, align=center, font=\footnotesize,},
	ytick = {2, 3, 4, 5, 6, 7, 8},
	legend columns=2,
	legend style={at={(0.02,0.24)},anchor=south west, font=\fontsize{6}{5}\selectfont, text width=1cm,text height=0.075cm,text depth=.ex, fill = none,},
	]
	
	\addplot[color=rred, mark = star, mark options = {scale = 0.8}, line width = 0.8pt, style = dashed] coordinates 
	{
		(10, 7.3627)
		(20, 7.2525)
		(30, 7.1609)
		(40, 7.1215)
		(50, 7.1014)
		(60, 7.0478)
		(70, 7.0106)
		(80, 6.9548)
		(90, 6.8992)
		(100, 6.7457)
	}; \addlegendentry{PA}
	
	\addplot[color=oorange, mark = star, mark options = {scale = 0.8}, line width = 0.8pt, style = dashed] coordinates 
	{
		(10, 7.3546)
		(20, 7.2520)
		(30, 7.1420)
		(40, 7.1025)
		(50, 7.0283)
		(60, 6.9738)
		(70, 6.8821)
		(80, 6.7579)
		(90, 6.5726)
		(100, 5.7599)
	}; \addlegendentry{GA}
	
	\addplot[color=ggreen, mark = star, mark options = {scale = 0.8}, line width = 0.8pt, style = dashed] coordinates 
	{
		(10, 3.1387)
		(20, 2.9843)
		(30, 2.7262)
		(40, 2.6347)
		(50, 2.5755)
		(60, 2.5115)
		(70, 2.3681)
		(80, 2.4221)
		(90, 2.3840)
		(100, 2.3453)
	}; \addlegendentry{RA}
	
	\addplot[color=black, mark = pentagon*, mark options = {scale = 1.5, fill = rred}, line width = 1pt] coordinates 
	{
		(10, 7.1726)           
		(20, 7.0910)     
		(30, 7.0348)    
		(40, 6.9953)   
		(50, 7.0034)    
		(60, 6.8997)    
		(70, 6.8595)    
		(80, 6.8337)    
		(90, 6.7849)
		(100, 6.6248)     
	}; \addlegendentry{PA - 4 bits}
	
	\addplot[color=black, mark = triangle*, mark options = {scale = 1.5, fill = oorange}, line width = 1pt] coordinates 
	{
		(10, 7.1729)           
		(20, 7.0827)     
		(30, 6.9885)    
		(40, 6.9788)   
		(50, 6.8976)    
		(60, 6.7791)    
		(70, 6.7061)    
		(80, 6.5526)    
		(90, 6.4426)
		(100, 5.5929)     
	}; \addlegendentry{GA - 4 bits}
	
	\addplot[color=black, mark = square*, mark options = {fill = ggreen}, line width = 1pt] coordinates 
	{
		(10, 3.1787)           
		(20, 2.9050)     
		(30, 2.7263)    
		(40, 2.6635)   
		(50, 2.6153)    
		(60, 2.4456)    
		(70, 2.4251)    
		(80, 2.3743)    
		(90, 2.3443)
		(100, 2.3092) 
	}; \addlegendentry{RA - 4 bits}
	\end{axis}
	\end{tikzpicture}
	\caption{Worst-rate vehicle (4 bits)}
	\label{f9}
	\vspace{-0.25cm}
\end{figure}
We presented a subchannel assignment approach for V2V  \textit{mode-3} communications based on weighted bipartite graph matching. We also considered relevant constraints that are to be enforced in order to avoid intra-cluster conflicts. Furthermore, the mentioned approach is compared against greedy and random algorithms. All three approaches were assessed using both fine-grained and quantized SINR values. When employing the proposed approach, as few as 2 quantization bits are enough for attaining efficient subchannel assignment. However, if the greedy algorithm is employed, 3 quantization bits will be required in order not to deviate notoriously from the ideal fine-grained curve performance. The type of quantization considered herein is linear.

\end{document}